# Evidence of unconventional superconductivity in the Ni-doped $NbB_2$ system

S. T. Renosto,[1] R. Lang,[2*] E. Diez,[1] L. E. Corrêa,[3] M. S. da Luz,[4] Z. Fisk,[5] and A. J. S. Machado[3]

[1] Universidad de Salamanca - Laboratorio de Bajas Temperaturas, Salamanca, 37008, Spain

[2] Universidade Federal de São Paulo - Instituto de Ciência e Tecnologia, São José dos Campos, 12231-280, Brazil

[3] Universidade de São Paulo - Departamento de Materiais, Lorena, 12602-810, Brazil

[4] Universidade Federal do Triângulo Mineiro, Uberaba, 38066-200, Brazil

[5] University of California at Irvine - Department of Physics and Astronomy, Irvine, CA 92697, USA

**Abstract**

In this paper, a comprehensive study of the effects of Ni-doping on structural, electrical, thermal and magnetic properties of the $NbB_2$ is presented. Low amounts ($\leq 10$ %) of Ni substitution on Nb sites cause structural distortions and induce drastic changes in the physical properties, such as the emergence of a bulk superconducting state with anomalous behaviors in the critical fields (lower and upper) and in the specific heat. Ni-doping at the 9 at.% level, for instance, is able to increase the critical temperature ($T_C$) in stoichiometric $NbB_2$ ($< 1.3$ K) to approximately 6.0 K. Bulk superconductivity is confirmed by magnetization, electronic transport, and specific heat measurements. Both $H_{c1}$ and $H_{c2}$ critical fields exhibit a linear dependence with reduced temperature ($T/T_C$), and the specific heat deviates remarkably from the conventional exponential temperature dependence of the single-band BCS theory. These findings suggest multiband superconductivity in the composition range from $0.01 \leq x \leq 0.10$ ($Nb_{1-x}Ni_xB_2$).

Keywords: Multiband superconductivity; $NbB_2$ system; Ni substitution; Linear critical fields.

*Corresponding author: rossano.lang@unifesp.br

## I. Introduction

Since the discovery of multiple gap structure (two-gap) of the superconducting state (below $\approx 40$ K) in $MgB_2$, the physical phenomenon of multiband superconductivity has received considerable attention [1-3]. Properties and characteristics of similar metal diborides have been extensively investigated, with the challenge of finding new superconducting materials which crystallize in an $AlB_2$-prototype structure ($P6/mmm$ space group). In fact, superconductivity has been reported in several metal diborides [4,5].

However, the origin of superconductivity of $NbB_2$ is still a matter of debate in the literature with the critical temperature ($T_C$) ranging from 4.6 K to 9.2 K; values usually determined by magnetization ($M$) [6-8]. Diamagnetic transition measurement can be confused by the magnetic shielding of bulk material and lead to a misinterpretation. This effect often arises from the existence of superconductor-metal segregation at grain boundaries, with a volume fraction lower than the sensitivity of experimental techniques such as X-ray diffraction. Even in zero-field resistivity ($\rho$) measurements segregation can provide percolative paths for superconducting current where proximity effects are included. Two features of NbB$_2$ suggest such possibilities: *i*) a large composition existence (NbB$_{1.92}$ to NbB$_{2.28}$) [9] that could cause unintentional deviations in composition making the AlB$_2$ majority structure even with metal segregation, and *ii*) the very high melting temperature (~ 3000 °C) which can promote B loss by sublimation during the synthesis and the subsequent remanence of liquid metal in the boundaries after primary solidification. Synthesis at temperatures lower than the melting point, for example by solid-state reaction from the mixture of Nb + B powders, can contain unreacted material [10].

Bulk measurements such as magneto-resistivity and specific heat are therefore necessaries to confirm the $T_C$ and the superconducting behavior for this compound. Several authors reported superconductivity reaching close to 9 K in NbB$_2$ single crystal [11]. However, these results have never been reproduced [12]. Nb contamination cannot be ruled out completely (Nb has a $T_C$ ~ 9.6 K), as recently reported for NbB [13]. There may be a large volumetric fraction of NbB$_2$, but also an amount of "almost pure" Nb that behaves as a superconducting material. Mudgel *et al.* inferred the absence of a superconductor state above 2.0 K in stoichiometric NbB$_2$ by $M(T)$ and $\rho(T)$ measurements [14]. Indeed, in 1966 Ukei and Kanda measured bulk superconductivity in stoichiometric NbB$_2$ with a transition below 1.0 K [15]. Thirteen years later, a critical temperature close to 0.62 K (from specific heat measurements) was confirmed by Leyarovska with the absence of any transition above to 1.0 K [16]. Nonetheless, some research groups claim that superconductivity only exists in Nb-deficient phases (Nb$_{1-x}$B$_2$) where $T_C$ is related to the Nb vacancy concentration [8,17]. Historically, superconductivity was found in non-stoichiometric NbB$_{1.94}$ at 1.2 K in 1951 [18]. These findings have then warranted further investigation in NbB$_2$. Nunes *et al.* reported a systematic study on the defect structure of NbB$_2$ by neutron diffraction [19]. The study revealed that for all the compositions of NbB$_{2-x}$, the maximum $T_C$ is lower than 3.5 K and the distance between Nb-B atoms remained approximately constant at 2.43 Å. This result suggests that the lattice parameters variation as a function of B content does not occur randomly. To explain the large stability interval within the NbB$_2$ phase, a structure of defects created by vacancies has been proposed [20]. The defective structure is based on cohesive forces in the planes of B, balanced by expansive forces in the planes of Nb. As a consequence of these defects, the electronic structure and the electron-phonon coupling are extremely dependent on the stoichiometry. No less important is the possibility that vacancies at both the Nb and B sites can receive unintentional contaminants or dopants that stabilize the stoichiometric and non-stoichiometric phases.

From these perspectives, we investigate the effects of Ni-doping on Nb sites in NbB$_2$. A thorough study reveals the stability of the AlB$_2$-prototype during the metallic substitution in Nb$_{1-x}$Ni$_x$B$_2$ ($0 \leq x \leq 0.10$ range). A dense set of magnetic, electrical and thermal measurements demonstrate that Ni-doping is able to increase the $T_C$ of stoichiometric NbB$_2$. In particular, specific heat results show clear evidence of unconventional superconductivity for the Nb$_{0.91}$Ni$_{0.09}$B$_2$ nominal composition with a critical temperature close to 6.0 K. For this composition, the lower and upper critical fields ($H_{c1}$ and $H_{c2}$) have a linear behavior and the specific heat ($C_p$) at very low temperatures deviates strongly from the conventional exponential temperature dependence. Such experimental results are beyond those addressed by single-band Bardeen-Cooper-Schrieffer (BCS) theory and suggest an anomalous superconducting behavior in Nb$_{1-x}$Ni$_x$B$_2$, possibly arising from multiband effects.

## II. Experimental Procedures

Polycrystalline $Nb_{1-x}Ni_xB_2$ samples in the composition range from $0 < x \leq 0.10$ were synthesized by arc melting. Stoichiometric amounts of high purity elements Nb, Ni, and B (> 99.99 %) were melted on a water-cooled Cu crucible using a non-consumable W electrode in a horizontal arc-furnace with very high electrical current (380 A ~ 3500 °C) and elevated heat extraction. After several high vacuum purges, the melting was carried out under high purity Argon atmosphere gettered by melting a Ti sponge. Samples of 0.25 g were flipped over and re-melted 5 times to ensure their good homogeneity. The weight loss during the arc melting was negligible (< 0.5 %). As already mentioned, stoichiometry deviations in $NbB_2$ can induce superconductivity and changes in the superconducting critical temperature. In order to conduct a detailed study of the Ni-doping effects in the $NbB_2$ system, mass losses were systematically controlled to keep a final composition with 1:2 of metal: boron atomic ratio.

X-ray powder diffraction (XRD) was obtained in a Panalytical diffractometer (model Empyrean) with detector PIXcel$^{3D}$ accessory using Cu-K$\alpha_1$ (1.5406 Å) and Bragg-Brentano geometry. The diffraction patterns were acquired in the 2θ range between 20° and 90° with steps of 0.05° and acquisition time of 2 s. The lattice parameters analyses, patterns simulation, and refinement of the structures (Rietveld method) were performed using PowderCell [21] and GSAS software [22] adopting as reference the $NbB_2$ phase crystallographic data reported in the literature [19].

Magnetic, electric, and thermal characterizations of as-cast samples were carried out by using a Quantum Design system: MPMS-SQUID and PPMS-9T with He$^3$ probe. The $T_C$ (onset) was defined from resistivity, magnetization and specific heat as the first temperature signal from normal to superconductor state. Magnetization ($M$) measurements were performed with vibrating sample measure system (VSM); a DC external field of 10 Oe with both zero field cooling (ZFC) and field cooling (FC) regimes in the temperature range from 2 to 10 K. The hysteresis loops of $M$ *versus* applied magnetic field ($H$) curves were acquired at 2 K in the $-3$ kOe $\leq H \leq 3$ kOe range. Electrical transport measurements were done by using the conventional four-point method, with a probe current of 1 mA. The $\rho$ *versus* $T$ curves were obtained in the temperature range from 2 to 300 K, and the magneto-resistivity was performed with applied $H$-field between $0 \leq H \leq 7$ kOe.

Specific heat ($C_p$) of polished flat $NbB_2$ and $Nb_{0.91}Ni_{0.09}B_2$ samples were measured with and without an applied field in the $0 \leq H \leq 7$ kOe range, between the temperatures from the 0.5 to 10 K using the relaxation method with calorimeter coupling to He$^3$ system (Quantum Design and Triton-Oxford).

## III. Results and Discussion

In the discussion that follow, we assume the Ni nominal content, i.e., that the samples stoichiometry equals the ratio of elements that went into the melt. Figure 1 shows the XRD patterns obtained for the $Nb_{1-x}Ni_xB_2$ samples with $x$ from 0 up to 0.10. For the entire set of samples, Fig 1(a), the (*hkl*) reflection peaks can be indexed using the AlB$_2$-prototype (*P6/mmm* space group) related to the $NbB_2$ phase. The 2θ range between 25° and 35°, related to the *a* and *c* axes of the hexagonal structure, is shown in detail in Fig. 1(b). The diffraction peaks shift to higher angles with increasing Ni-doping, indicating unit cell contraction. Rietveld refinement analysis confirms that all samples crystallize in the hexagonal $AlB_2$ structure. From the results were extracted values of lattice parameter, unit cell volume, refinement *R*-factors and goodness of fit (*S*) for each sample. Table I shows the obtained values. Figure 1(c) depicts in particular the refinement for the $Nb_{0.91}Ni_{0.09}B_2$ sample, i.e. with $x = 0.09$. The refined *a* and *c* lattice parameters are 3.0869(4) and 3.2226(3) Å, respectively. It are slightly lower than that observed for the stoichiometric $NbB_2$, where $a = 3.1125(5)$ Å and $c = 3.2653(4)$ Å. This points to the dependence expected upon Ni substitution since the Ni atomic radius (124 pm) is smaller than that of Nb (146 pm).

**Table I.** From Rietveld refinement: lattice parameter, unit cell volume, *R*-factors ($R_{wp}$, $R_p$, and $R_e$) and goodness of fit (*S*) for each sample.

| Ni nominal content [%] | Lattice parameter (*a*) [Å] | Lattice parameter (*c*) [Å] | Unit cell volume [Å$^3$] | $R_{wp}$ [%] | $R_p$ [%] | $R_e$ [%] | *S* |
|---|---|---|---|---|---|---|---|
| 0 | 3.1125(5) | 3.2653(4) | 27.395(6) | 9.73 | 7.13 | 4.74 | 4.21 |
| 0.01 | 3.1042(5) | 3.2475(4) | 27.101(7) | 11.56 | 8.52 | 5.12 | 5.09 |
| 0.03 | 3.0991(4) | 3.2379(3) | 26.932(6) | 13.43 | 10.1 | 6.36 | 4.45 |
| 0.05 | 3.0956(5) | 3.2272(4) | 26.782(6) | 12.24 | 7.43 | 4.22 | 8.41 |
| 0.07 | 3.0897(4) | 3.2254(3) | 26.665(7) | 11.47 | 9.78 | 5.67 | 4.09 |
| 0.09 | 3.0869(4) | 3.2226(3) | 26.594(6) | 12.37 | 8.75 | 4.34 | 8.12 |
| 0.10 | 3.0882(4) | 3.2237(3) | 26.625(6) | 15.89 | 10.65 | 6.23 | 6.50 |

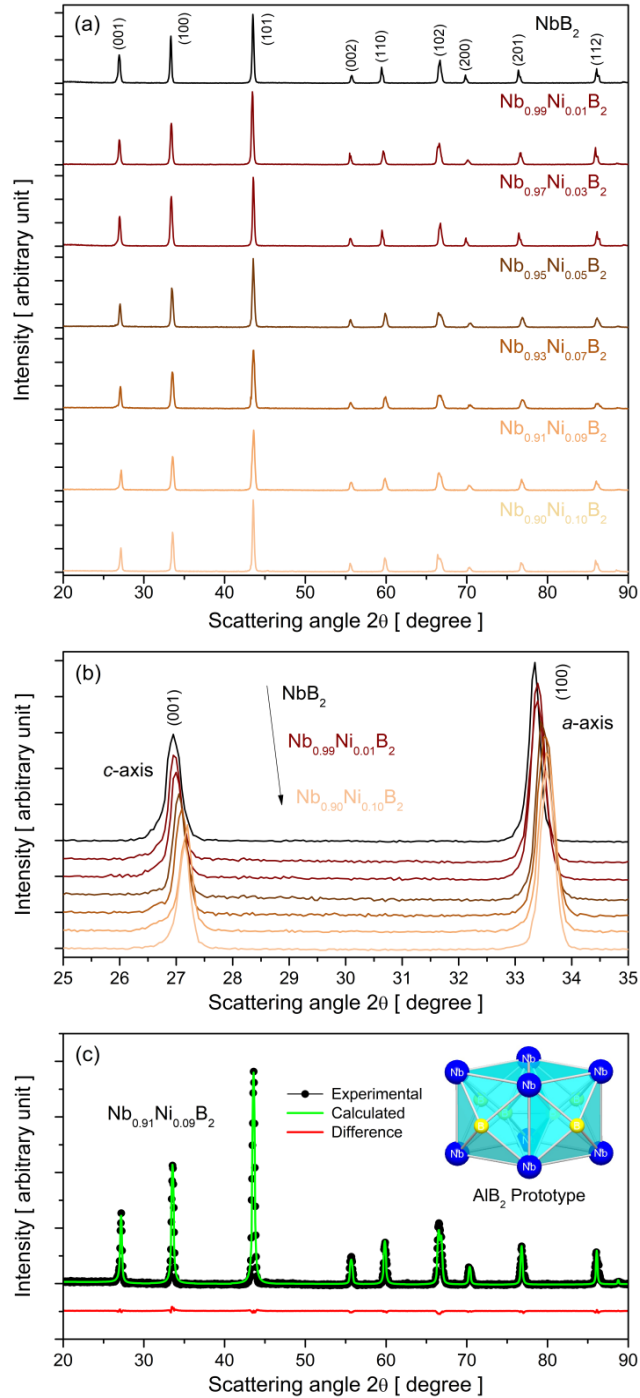

**FIG. 1. (a)** X-ray-diffraction patterns of samples with $Nb_{1-x}Ni_xB_2$ nominal composition from $0 \leq x \leq 0.10$. The (*hkl*) Muller indices denote the reflections from the $NbB_2$. **(b)** Diffraction range related to the *a* and *c* axes of the hexagonal structure in detail. **(c)** Fitting result of the XRD Rietveld refinement for the $Nb_{0.91}Ni_{0.09}B_2$ sample. The inset illustrates the hexagonal crystalline structure where the blue spheres represent the metal and yellow the B atoms.

The dependence of the lattice parameters on Ni substitution reveals a sensitive and small contraction of the unit cell (volume) with increasing Ni in the host matrix (partial replacement of Nb sites) as observed in Figs.

1(b) and 2(a). Interestingly, the contraction appears to occur nonuniformly along the different crystal axes, since the unit cell is anisotropic. According to the refinement results, there is a greater variation of the *c*-axis (1.31 %) when compared with the basal parameter (0.82 %). Under closer inspection, the diffraction data for composition higher than $x = 0.09$ suggest a saturation of the lattice parameters contraction, i.e., an indication of the existence of a solubility limit for Ni substitution on the Nb atomic sites, very close to this composition. Indeed, a low solubility limit is consistent with the absence of a $NiB_2$ phase in the Ni-B phase diagram. The ionic radius difference between Ni and B ions is approximately 27 %, pointing out that a complete solid solution cannot be formed in this system. Nonetheless, we cannot rule out the possibility of inhomogeneity of the samples.

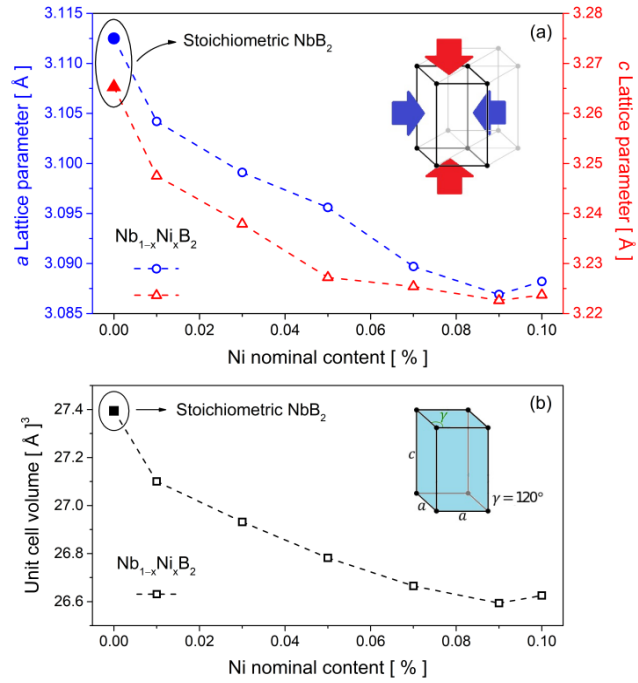

**FIG. 2. (a)** Lattice parameters and **(b)** unit cell volume as a function of Ni concentration in $Nb_{1-x}Ni_xB_2$ samples in the range from $0 \leq x \leq 0.10$.

Figure 3 shows the *M(T)* measurements of the $Nb_{1-x}Ni_xB_2$ samples in ZFC and FC regimes (DC external field of 10 Oe) in the temperature range from 2 to 10 K. The stoichiometric $NbB_2$ curve is also shown for comparison purposes. No transition and hysteresis are detected for $NbB_2$ sample above 2 K, whereas for the $Nb_{1-x}Ni_xB_2$ samples, clear diamagnetic transitions between 3.9 and 6.0 K are observed, evidence that the superconductivity in this diboride is strongly dependent on the metal substitution. The difference between the FC and ZFC signals is mainly due to the imprisonment of vortices by grain boundaries and point magnetic moments in superconducting material. In the ZFC regime below $T_C$ there is a finite potential barrier around the vortices, and in this case, the magnetic flux is pinned in the bulk of the material. This drastically reduces the magnetization in the FC regime.

The *M(H)* dependence exhibited in the insets (at 2 K in the $-3$ kOe $\leq H \leq 3$ kOe range), suggest II-type superconducting behavior. In addition, the shielding in the Meissner state allows estimating a superconducting fraction (*SF*) of up to 90 % for compositions below $Nb_{0.91}Ni_{0.09}B_2$. It is tempting to suppose that Ni insertion drastically affects the $NbB_2$ electronic structure, inducing bulk superconductivity above 2 K, since a transition has emerged even at a very low level of Ni substitution ($x = 0.01$). However, for the $Nb_{0.90}Ni_{0.10}B_2$ composition the critical temperature decreases slightly from a maximum, where $T_C \approx 5.3$ K. Actually, the normalized magnetic moment value (emu/g) at 2 K also has reduced, suggesting an *SF* decrease. Both $T_C$ and *SF* decreases can be due to magnetic pair breaking effects by possible Ni segregation to this substitution level ($x = 0.1$).

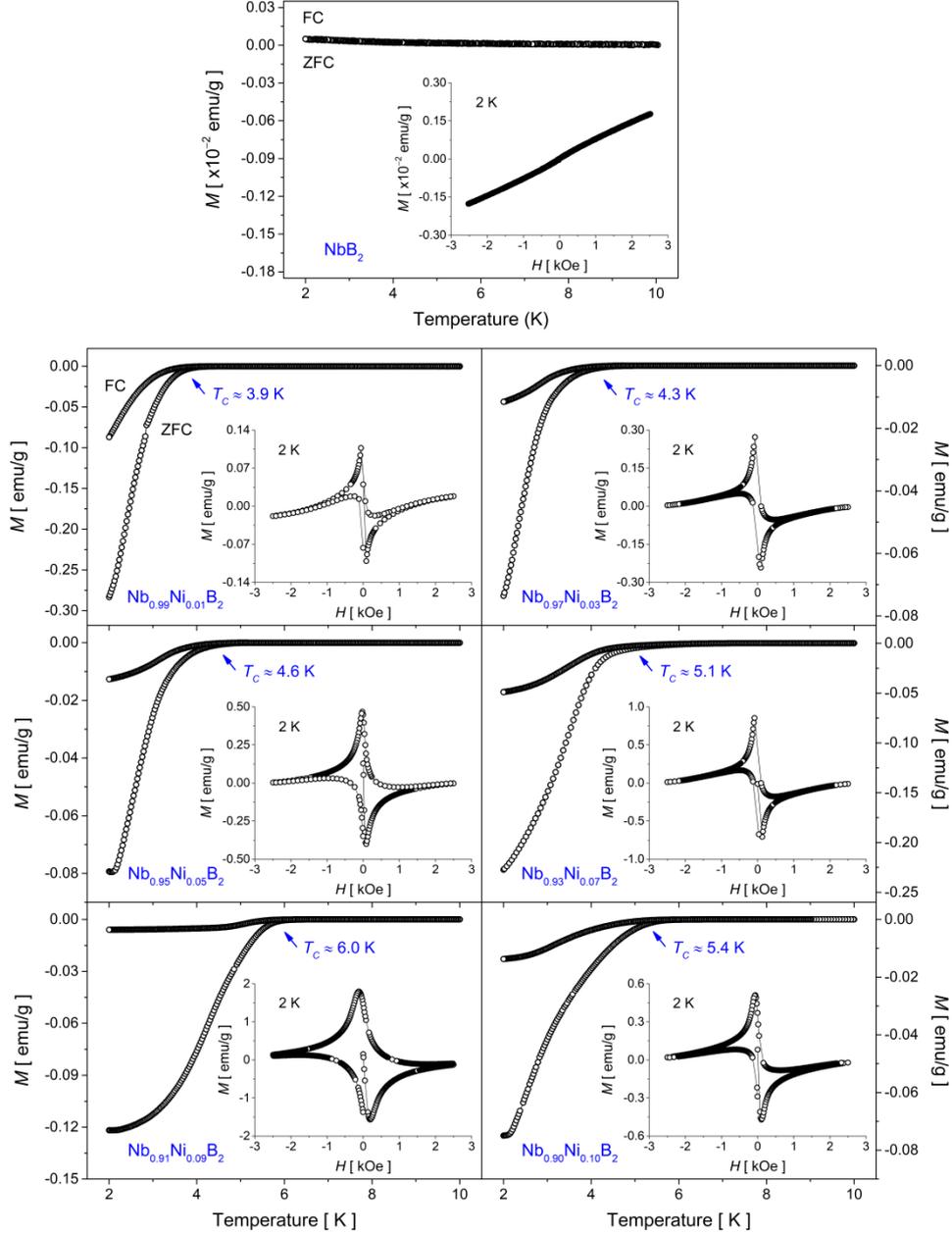

**FIG 3.** Temperature dependence of the magnetization of the $Nb_{1-x}Ni_xB_2$ samples in the range from $0 \leq x \leq 0.10$. The insets show the magnetization (*M*) as a function of the applied magnetic field (*H*) at *T* = 2 K.

Figure 4 presents a summary of the magnetization results that shows $T_C$ as a function of the Ni nominal content ($x$). Critical temperature increases with an increase in Ni-doping reaching a maximum at $x = 0.09$. For the $Nb_{0.90}Ni_{0.10}B_2$ composition, $T_C$ goes to a lower value. This result may be associated with the solubility limit for Ni substitution on Nb sites or a sample inhomogeneity as suggested by the X-ray diffraction findings. Although further studies should be conducted to clarify this aspect, we assume that the Ni substitution level for optimized superconductivity in the range of composition investigated is $x = 0.09$.

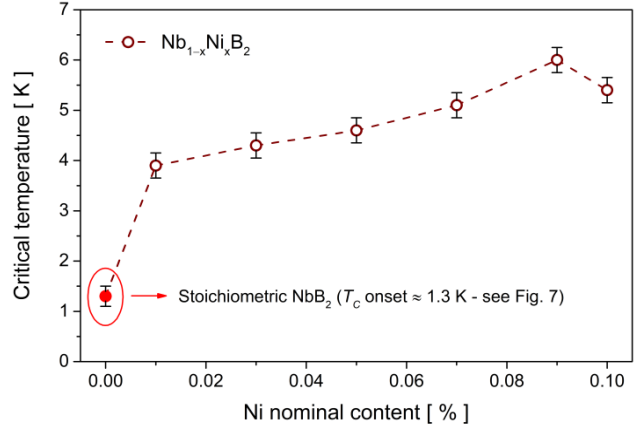

**FIG 4.** Critical temperature ($T_C$) as a function of Ni-doping in the $NbB_2$ system. The dashed line is a guide for the eyes.

Figure 5 shows electrical resistivity measurements as a function of temperature (2 - 300 K) and as a function of applied magnetic field (0 - 7 kOe) of the $Nb_{0.91}Ni_{0.09}B_2$ composition that is believed to be the optimum $T_C$. The $\rho(T)$ curve, Fig. 5(a), exhibits a sharp superconducting transition close to 6.5 K (better seen in the inset). Such a feature corroborates the magnetization ones (Fig. 3). Furthermore, shifts of $T_C$ (typical of superconductor character) can be observed in the normalized magneto-resistivity curves as shown in the Fig. 5(b). Undoubtedly, $\rho \to 0$ in applied $H$-field supports the existence of a bulk superconducting state in the sample.

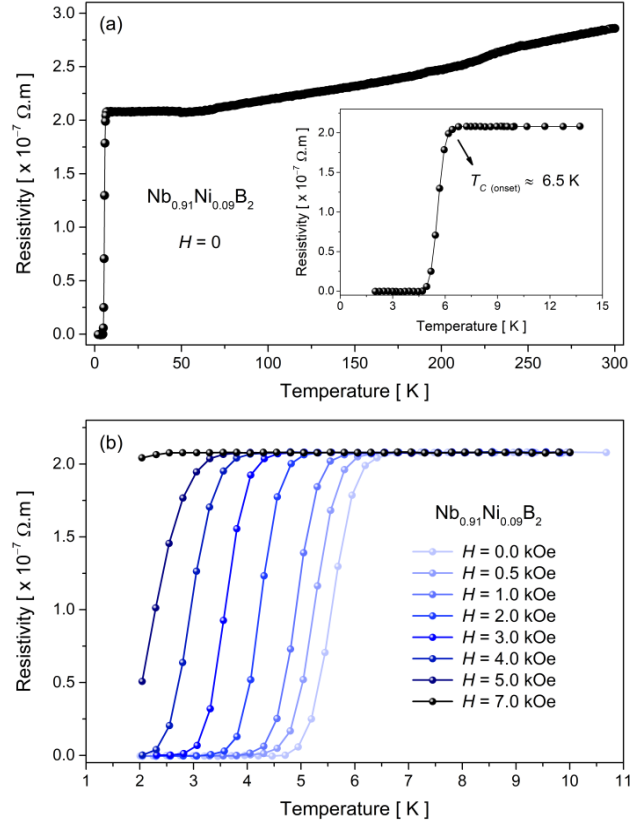

**FIG. 5.** Temperature dependence of the **(a)** resistivity and **(b)** magneto-resistivity of the $Nb_{0.91}Ni_{0.09}B_2$ sample.

A detailed investigation of the lower critical field ($H_{c1}$) for $Nb_{0.91}Ni_{0.09}B_2$ is presented in Fig. 6. Magnetization curves in the temperature range between 1.8 and 6.0 K in applied $H$-fields up to 600 Oe (Fig. 6(a)), nearly to the Meissner state, were obtained using a rectangular bar-shaped sample in order to minimize the demagnetization factor. The $H_{c1}$ values were then estimated from the deviation point of the linear slope of the curves and by assuming $\Delta M = 1 \times 10^{-3}$ emu/g as a criterion to distinguish the Meissner line from the $M$ signal [23] as shown in Fig. 6(b). Once the $H_{c1}$ and $T_C$ data are obtained (from the Figs. 6(b) and 3, respectively), the $H_{c1}$ behavior with the reduced temperature ($\tau = T/T_C$) can be evaluated. Figure 6(c) shows the range of experimental data where a linear dependence is established. An $H_{c1}$ approximately 116 Oe is obtained assuming a linear extrapolation up to $\tau = 0$ K. However, for conventional superconducting materials, the lower critical field as a function of the reduced temperature can be described in terms of an empirical expression with a quadratic dependence:

$$H_{c1}(\tau) = H_{c1(0)}\left(1 - \tau^2\right), \tag{1}$$

where $H_{c1(0)}$ is the lower critical field at zero Kelvin. The $H_{c1}(\tau)$ curve is shown for a comparison approach, admitting experimental data directly on the Eq. (1) with no restrictions on $H_{c1(0)}$. Such a quadratic dependence does not seem to be observed and therefore provides evidence of unconventional superconductivity in the $Nb_{1-x}Ni_xB_2$ system. Similar features were also found in $MgB_2$ and $Zr_{0.96}V_{0.04}B_2$, which do not seem to follow the expected trend of the single-band BCS theory [4,23-26].

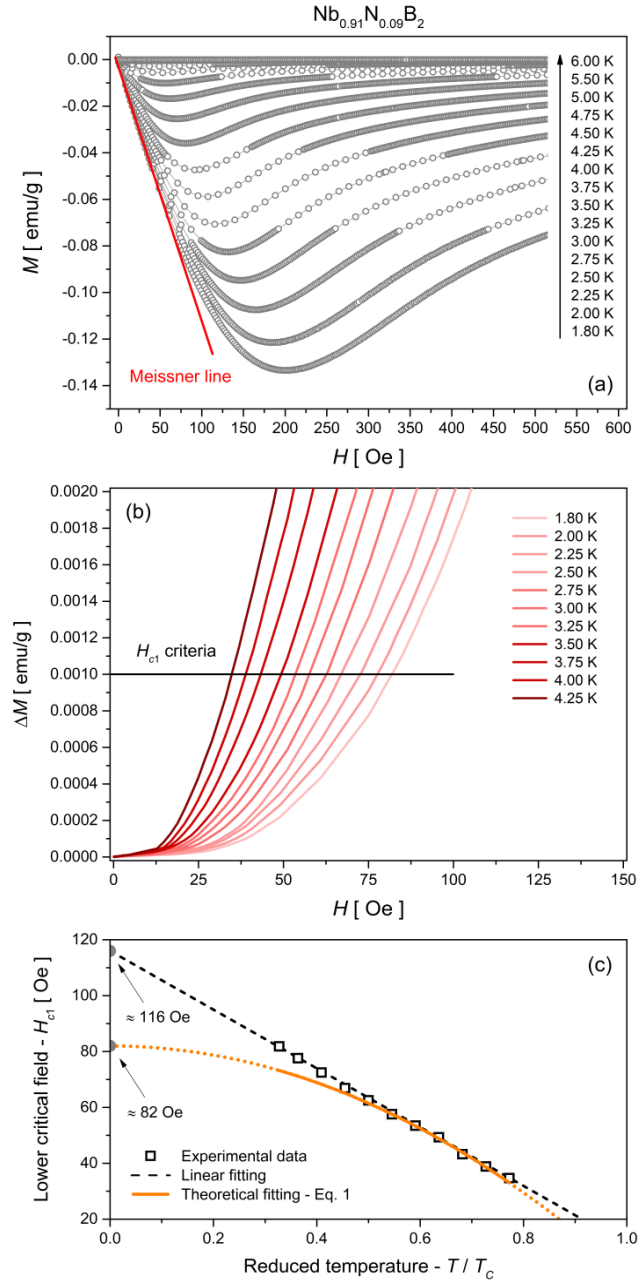

**FIG. 6.** $Nb_{0.91}Ni_{0.09}B_2$ sample: **(a)** applied magnetic field ($H$) dependence of magnetization ($M$) for different temperatures ($T$). **(b)** $\Delta M$ *versus* $H$ used as a criterion for $H_{c1}$ definition. **(c)** Lower critical field ($H_{c1}$) *versus* reduced temperature ($\tau = T/T_C$).

For a complete understanding of the observed behavior, specific heat ($C_p$) measurements at very low temperatures were made. Figures 7(a) and 7(b) show $C_p/T$ versus $T^2$ in several applied $H$-fields (0 - 7 kOe) for $NbB_2$ and $Nb_{0.91}Ni_{0.09}B_2$, respectively. A well-defined anomaly (jump) at 0.8 K (onset at $\approx$ 1.3 K) is observed in the $NbB_2$ curve at zero external field (Fig. 7(a)) and corresponds to the phase transition from the normal (non-superconducting) to the superconducting state, in line with the findings [15,16]. In contrast, the anomaly is stronger and broadened close to 6.0 K at $H = 0$ for the Ni-containing sample, Fig. 7(b). The standard procedure is to find a sharp and discontinuous jump for a conclusive occurrence of a bulk second order transition with a strong break in the $C_p/T$ linearity, however, the superconducting transition is much more spread out that would be expected and therefore we take the maximum rather than onset. Nevertheless, the $T_C$ value of ≈ 4.0 K (onset at ≈ 6.0 K) agrees well with the resistivity and magnetization results where an equivalent superconducting critical temperature is seen (Figs. 5 and 4). In addition, the shifts of $T_C$ to lower temperatures with the increase of the applied field are consistent with the existence of a superconducting state.

The specific heat was analyzed by fitting in normal-state from 10 K up to near the $T_C$ following Debye model:

$$C_{p\ (T\to 0)} = \gamma T + \beta T^3. \qquad (2)$$

The first part of the expression corresponds the electronic contribution to the $C_p$ described by the Sommerfeld coefficient $\gamma = 1/3[\pi^2 k_B^2(1+\lambda_{EP})]$DOS, where $k_B$ is the Boltzmann's constant and DOS the total density of states at the Fermi level. The parabolic dispersion is corrected by the electron-phonon coupling term $(1+\lambda_{EP})$ that renormalizes the electronic specific heat. The second part is related to the phononic contribution described by $\beta = 12/5(N_A k_B \pi^4 \Theta_D^{-3})$, where $N_A$ is the Avogadro's number, and $\Theta_D$ is the Debye characteristic temperature. With a least-square fit, both $\gamma$ and $\beta$ values were evaluated for the NbB$_2$ and Nb$_{0.91}$Ni$_{0.09}$B$_2$ samples. For NbB$_2$, $\gamma \approx 2.217$ mJ/mol·K$^2$, $\beta \approx 0.022$ (mJ/mol·K$^4$) and $\Theta_D \approx 642$ K. The $\gamma$ value agrees well with those reported in the literature ($\gamma \approx 2.33$ mJ/mol·K$^2$) for AlB$_2$-type transition-metal diborides [27]. The estimated $\Theta_D$ value by means of the $\beta$ term is also close to that calculated for NbB$_2$ using a rigid ion model (RIM): $\Theta_D \approx 739$ K [28]. For Nb$_{0.91}$Ni$_{0.09}$B$_2$ at zero field, $\gamma \approx 3.602$ (mJ/mol·K$^2$), $\beta \approx 0.046$ (mJ/mol·K$^4$) and $\Theta_D \approx 500$ K. In addition, a subtle dependence of $\gamma$ with applied field is observed, where $\gamma$ increases with the applied field while $\beta$ remains constant. However, at zero field, both $\gamma$ and $\Theta_D$ are significantly different when compared with the NbB$_2$ data. The electronic contribution increases whereas the Debye temperature decreases, which indicates changes in the total density of states and in the phonon spectrum (it will be better discussed in what follows) for the Ni-containing sample.

An average density of states <DOS> can be assessed knowing the $(1+ <\lambda_{EP}>)$ term. The electron-phonon coupling constant $\lambda_{EP}$ can be determined from $T_C$ transition temperature and the $\Theta_D$ Debye temperature by the following McMillan expression [29]:

$$\lambda_{EP} = \frac{\mu^* \ln(\Theta_D/1.45\,T_C) + 1.04}{(1-0.62\,\mu^*)\ln(\Theta_D/1.45\,T_C) - 1.04}. \qquad (3)$$

This description has been based on Eliashberg equations with microscope elements (mainly the electron-phonon spectral function) in the BCS theory. Although the Eq. (3) has been derived for an isotropic Fermi surface, several aspects of theoretical fundaments agreed well with an anisotropic superconducting where the determination of a <$\lambda_{EP}$> from an observable, such as the specific heat is still valid [30]. Therefore, we use Eq. (3) to evaluate the impact of Ni-doping on the NbB$_2$ electronic property. The effective pseudopotential repulsion ($\mu^*$) that arises from the propagation difference between the Coulomb and phononic couplings in the vicinity of $T_C$ was considered $\mu^* \approx 0.1$ [31,32].

For NbB$_2$, the <$\lambda_{EP}$> value is about 0.36, consistent with the observed low $T_C \approx 0.8$ K, and indicates a weak coupling in the BCS scenario. Indeed, calculations of the total coupling constant $\lambda$ ($\lambda_{tot} = \lambda_{Nb} + 2\lambda_B$) for stoichiometric NbB$_2$ suggest a <$\lambda_{EP}$> ≈ 0.387, with an expressive B atoms contribution ($\lambda_B \approx 0.316$) [33].

Substituting (1+ <$\lambda_{EP}$>) as an average correction factor at the γ electronic coefficient expression, a <DOS> of ≈ 0.22 eV$^{-1}$ atom$^{-1}$ is obtained. For Nb$_{0.91}$Ni$_{0.09}$B$_2$, the average coupling constant is approximately 0.58, and consequently a <DOS> ≈ 0.32 eV$^{-1}$ atom$^{-1}$.

Therefore, the observations indicate an increase in the density of states (by a factor of 1.45) and a reduction of $\Theta_D$ (by a factor of 1.28, probably by a relaxation in the phonon spectrum) for the Ni-containing sample. A partial conclusion is that the Ni-doping not only introduces lattice distortions but also modifies the electronic/phonon structure of the precursor diboride. The effect of lower Ni mass (when compared to N$_B$) is clear and suggests that most of the coupling arises from the higher frequency vibration of the substituted Ni atoms at the Nb sites [34]. Furthermore, the lattice distortions due to the Ni-doping, particularly the compressive strains on the *c*-axis, could induce dimerization of semi-filled $p_z$-$p_z$ orbitals by approximation between B atomic layers with further influence on the electron-phonon coupling.

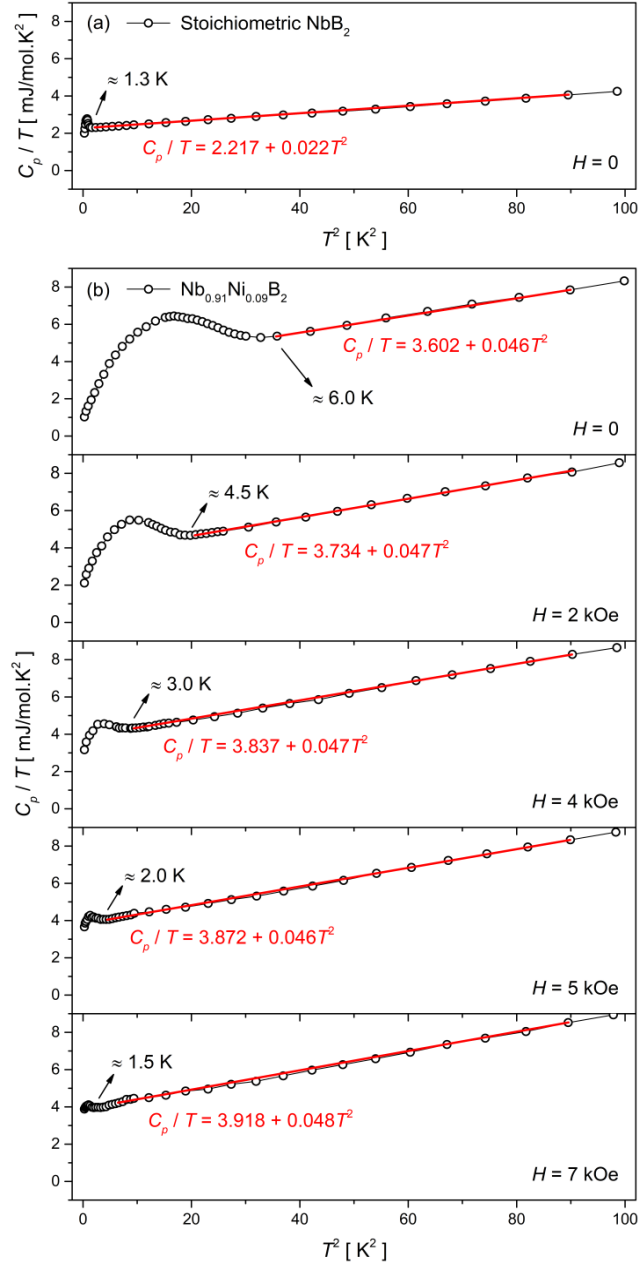

**FIG. 7.** Temperature dependence of the specific heat for several applied *H*-field. **(a)** for $NbB_2$ and **(b)** for $Nb_{0.91}Ni_{0.09}B_2$. The continuous line is a fitting based on the normal state contribution of the specific heat.

Figure 8(a) shows the applied *H*-field dependence of the anomaly size determined from the $C_{p(electronic)}/\gamma$ ratio, i.e., the $C_{p(electronic)} = C_p/T - \beta T^2$ electronic contribution by $\gamma$ in the normal state. At zero external field, before the transition $C_{p(electronic)}/\gamma = 1.0$ and at transition, the anomaly size ($\Delta C_{p(electronic)}/\gamma T_C$) is about 0.39 which is considerably smaller than that of the bulk pairing BCS prediction (1.43) [16]. On the other hand, the linearized electronic contribution $\ln(C_{p(electronic)}/\gamma T_C)$ to the $C_p$ *versus* $T_C/T$ displays a remarkable divergence from BCS theory up to close to $T_C/T \approx 2.8$ (Fig. 8(b)). BCS predicts that electron excitations with energies close to the Fermi level through the single isotropic gap gives rise to an exponential behavior at temperatures lower than $T_C$. On the other hand, in multiband material the increase of the electron-phonon coupling anisotropy leads to an non-exponential dependence [35,36]. As a first approach to describe the dependence on temperature, a power law fit was used as shown in Fig. 8(b). However, a complete explanation of the problem requires further calculations. In Fig. 8(c), the *H*-field dependence of the γ electronic coefficient can be observed, where γ values are the fits from the Fig. 7 curves and normalized by their value at $H = 7$ kOe. A similar result has been found for the $MgB_2$ multiband superconductor compound [37,38]. Pribulova *et al.* proposed that such dependence emerges as a consequence of the contribution of two different gaps on the Fermi surface [39]. In the multiband case, different Fermi velocities and gap asymmetries with different scattering rates and different DOS contributions result in anomalous superconductor behavior in applied magnetic field; the two gaps being connected by a coupling term which results in only one $T_C$ [30,35]. In the present study, the experimental results appear to be consistent and suggest that the mechanism responsible for the superconducting state does not follow the BCS theory of a single isotropic band. However, the specific heat measurements were not able to separate the contribution of the two bands [40].

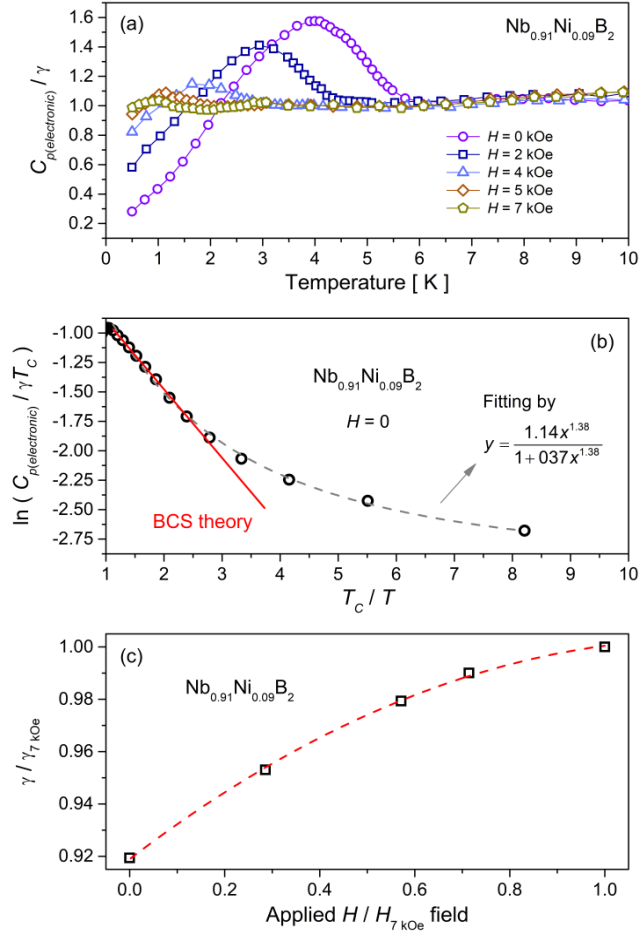

**FIG. 8.** (a) Normalized electronic contribution for the $Nb_{0.91}Ni_{0.09}B_2$ sample. (b) Linearized electronic contribution $\ln(C_{p(electronic)}/\gamma T_C)$ to the $C_p$ as a function of $T_C/T$, where the continuous line is a fitting based on the BCS theory. The dashed line is a nonlinear curve fitting where the temperature dependence of the data appears to be closer to a power law. **(c)** Applied $H/H_{7kOe}$ field dependence of the $\gamma/\gamma_{7kOe}$ electronic contribution. Here, the dashed line is only a guide by the eyes.

Figure 9 shows the upper critical field ($H_{c2}$) as a function of the reduced temperature ($\tau = T/T_C$). The data were extracted from the resistivity and specific heat measurements. The upper critical field at zero Kelvin ($H_{c2(0)}$), the point indicated in the figure, it was estimated by using the Werthamer-Helfand-Hohenberg (WHH) theory [41] in the limit of a short electronic mean-free path (dirty limit) given by: $H_{c2(0)} = -6.93T_C\,(dH_{c2}/dT)_{T\rightarrow Tc}$ (CGS unit). The derivative close to $T_C$ gives an $H_{c2(0)} \approx 8.2$ kOe. One observes in the superconducting phase diagram a striking linear behavior which differs from the second-order expected for single-band materials. Compounds considered multiband superconductors exhibit analogous response for $H_{c2}(\tau)$ [42-44].

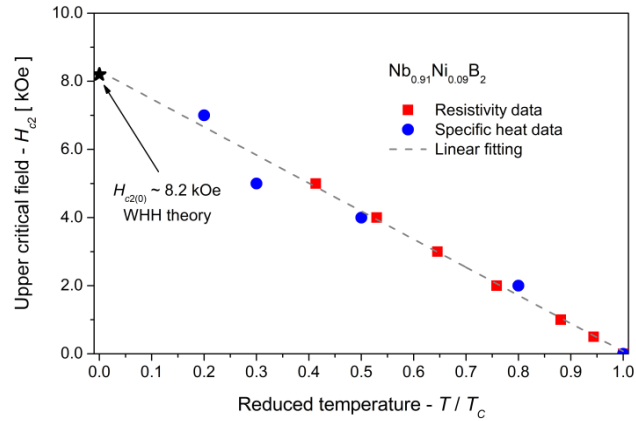

**FIG. 9.** Reduced temperature dependence of the upper critical field of the $Nb_{0.91}Ni_{0.09}B_2$ sample. The data were extracted from the resistivity and specific heat measurements. The dashed line corresponds to a linear fitting.

Doping with a magnetic impurity such as Ni has been investigated in several superconducting families [45-51]. Generally, Ni-doping suppresses the superconductivity and decreases the $T_C$. Magnetic pair breaking effects and changes in DOS near the Fermi level has been arguments accepted as a valid explanation for the outcome. Our results shows the opposite, where Ni-doping is able to increase the $T_C$ of $NbB_2$. Although the unconventional behavior origin observed in both critical fields and specific heat is not entirely clear, there are strong indications of multiband effects which seem to be typical of the $AlB_2$ prototype structure. Different scattering rates (interband and intraband) cannot be ruled out, as occurs in two-band superconductors. In particular, intraband dispersion by magnetic scattering (ferromagnetic Ni-doping) can affect the $H_{c2}$ critical field curvature without major influence on pair breaking effects. However, further investigations need to be conducted to clarify the mechanism which generates this superconductivity. Important questions about the nature of the substitutional doping effects on the electronic structure of $NbB_2$ such as Ni multiple valence states in association with the uncertainty of the substitution position becomes a significant issue for a theoretical description of the $Nb_{1-x}Ni_xB_2$ system.

**IV. Conclusions**

A detailed study of the superconductivity of polycrystalline $NbB_2$ and $Nb_{1-x}Ni_xB_2$ samples was presented and discussed. Specific heat results show that partial substitution of Nb by Ni ions generates strong changes in electronic and phononic structure, able increase the bulk superconductivity of the $NbB_2$ < 1.3 K to maximum $T_C \approx 6.0$ K at $Nb_{0.91}Ni_{0.09}B_2$ nominal composition. Excellent agreements for the transition temperature as characterized by magnetic, electrical and thermal measurements were observed. The lower and upper critical fields exhibit linear behaviors, in addition to the heat capacity data deviate from simple BCS prediction (exponential behavior), which we believe to indicate unconventional superconductivity possibly related to the multiband phenomenon.

**Acknowledgments**

The authors would like to thank the Brazilian funding agencies: FAPESP (2014/25235-3, 2018/08819-2), CNPq (300821/2012-3, 448041/2014-6), CAPES (CAPES/CNPq PVE A10/2013) and MINECO (MAT 2016/75955) for partial financial support. The authors are also gratefully acknowledged to Brazilian National Nanotechnology Laboratory (Proj. 12863 and 13555) and to Antônio L. R. Manesco for the discussions.

**References**

[1] J. Nagamatsu, N. Nakagawa, T. Muranaka, Y. Zenitani, and J. Akimitsu, Nature, **410**, 63 (2001).

[2] S. Souma, Y. Machida, T. Sato, T. Takahashi, H. Matsui, S.-C. Wang, H. Ding, A. Kaminski, J. C. Campuzano, S. Sasaki, and K. Kadowaki, Nature, **423**, 65 (2003).

[3] K.Chen, W. Dai, C. G. Zhuang, Q. Li, S. Carabello, J. G. Lambert, J. T. Mlack, R. C. Ramos, and X. X. Xi, Nature Comm. **3**, 619 (2011).

[4] S. T. Renosto, H. Consoline, C. A. M. dos Santos, J. Albino Aguiar, Soon-Gil Jung, J. Vanacken, V. V. Moshchalkov, Z. Fisk, and A. J. S. Machado, Phys. Rev. B, **87**, 174502 (2013).

[5] L. E. Muzzy, M. Avdeev, G. Lawes, M.K. Haas, H.W. Zandbergen, A. P. Ramirez, J. D. Jorgensen, and R. J. Cava, Physica C, **382**, 153 (2002).

[6] H. Takagiwa, E. Nishibori, N. Okadab, M. Takata, M. Sakata, and J. Akimitsu, Sci. Technol. Adv. Mater., **7**, 22 (2006).

[7] Z.-A. Ren, S. Kuroiwa, Y. Tomita, and J. Akimitsu, Physica C, **468**, 411 (2008).

[8] A. Yamamoto, C. Takao, T. Masui, M. Izumi, and S. Tajima, Physica C, **383**, 97 (2002).

[9] T. B. Massalski, H. Okamoto, P. R. Subramanian, and L. Kacprzak in *Binary Alloy Phase Diagrams*, 2nd ed., Metals Park: American Society For Metals (1990).

[10] T. Matsudaira, H. Itoh, S. Naka, H. Hamamoto, J. Less-Common Met., **155**, 207 (1989).

[11] J. E. Schirber, D. L. Overmyer, B.Morosin, E. L. Venturini, R. Baughman, D. Emin, H. Klesnar, and T. Aselage, Phys. Rev. B, **45**, 10787 (1992).

[12] T. Takahashi, S. Kawamata, S. Noguchi, and T. Ishida, Physica C **426**, 478 (2005).

[13] F. Abud, L. E. Correa, I. R. Souza Filho, A. J. S. Machado, M. S. Torikachvili, and R. F. Jardim, Phys. Rev. Materials, **1**, 044803 (2017).

[14] M. Mudgel, V. P. S. Awana, H. Kishan, I. Felner, Dr. G. A. Alvarez, and G. L. Bhalla, J. Appl. Phys., **105**, 07E313 (2009).

[15] K. Ukei and E. Kanda, Sci. Rep. Res. Inst., Tohoku Univ., Ser. **18** Suppl., 413 (1966). http://hdl.handle.net/10097/00102613.

[16] L. Leyarovska, and E. Leyarovski, J. Less-Common Met., **67**, 249 (1979).

[17] A. S. Cooper, E. Corenzwit, L. D. Longinotti, B. T. Matthias, and W. H. Zachariasen, Proceedings of the National Academy of Sciences, **67**, 313 (1970).

[18] J. K. Hulm, and B. T. Matthias, Phys. Rev., **82**, 273 (1951).

[19] C. A. Nunes, D. Kaczorowski, P. Rogl, M. R. Baldissera, P. A Suzuki, G. C. Coelho, A. Grytsiv, G. André, F. Boureé, and S. Okada, Acta Materialia, **53**, 3687 (2005).

[20] G. A. Meerson, and G.V Samsonov, J. Appl. Chem. USSR, **27**, 1053(1954).

[21] W. Kraus and G. Nolze, *Powder Cell for Windows*, http://www.ccp14.ac.uk/ccp/web-mirrors/powdcell/a_v/v_1/powder/e_cell.html, Berlin, version 2.3 (2000).

[22] A. C. Larson, and R. B. Von Dreele, *General Structure Analysis System* (*GSAS*), Los Alamos National Laboratory Report No. LAUR 86-748 (2004) (unpublished).

[23] S. L. Li, H. H. Wen, Z. W. Zhao, Y. M. Ni, Z. A. Ren, G. C. Che, H. P. Yang, Z. Y. Liu, and Z. X. Zhao, Phys. Rev. B, **64**, 094522 (2001).

[24] A. Gurevich, Physica C, 456, 160 (2007).

[25] I. N. Askerzade, A Gencer, and N. Güçlü, Supercond. Sci. Technol., **15,** L13 (2002).

[26] H. J. Choi, D. Roundy, H. Sun, M. L. Cohen, and S. G. Louie, Nature, **418**, 758 (2002).

[27] P. Vajeeston, P. Ravindran, C. Ravi, and R. Asokamani, Phys. Rev. B, **63**, 045115 (2001).

[28] N. Kaur , R. Mohan, N. K. Gaur, and R. K. Singh, Physica B **404**, 1607 (2009).

[29] W. L. McMillan, Phys. Rev., **167**, 331 (1968).

[30] H. J. Choi, D. Roundy, H. Sun, M. L. Cohen, and S. G. Louie, Phys. Rev. B, **66**, 020513(R) (2002).

[31] D. J. Scalapino, J. R. Schrieffer, and J. W. Wilkins, Phys. Rev., **148**, 263 (1966).

[32] R. C. Dynes, Solid State Commun., **10**, 615 (1972).

[33] P. J. T. Joseph, and P. P. Singh, Physica C, **391**, 125 (2003).

[34] E. Deligoz, K. Colakoglu, and Y.O. Ciftci, Solid State Commun., **150**, 405 (2010).

[35] D. Markowitz, and L. P. Kadanoff, Phys. Rev., **131**, 563 (1963).

[36] M. Zehetmayer, Supercond. Sci. Technol., **26**, 043001 (2013).

[37] H. D. Yang, J.-Y. Lin, H. H. Li, F. H. Hsu, C. J. Liu, S.-C. Li, R.-C. Yu, and C.-Q. Jin, Phys. Rev. Lett., **87**, 167003 (2001).

[38] Z. Pribulova, T. Klein, J. Marcus, C. Marcenat, F. Levy, M. S. Park, H.-G. Lee, B. W. Kang, S.-I. Lee, S. Tajima, and S. Lee, Phys. Rev. Lett., **98**, 137001 (2007).

[39] Z. Pribulova, T. Klein, J. Marcus, C. Marcenat, M. S. Park, H.-S. Lee, H.-G. Lee, and S.-I. Lee, Phys. Rev. B, **76**, 180502(R) (2007).

[40] F. Bouquet, R. A. Fisher, N. E. Phillips, D. G. Hinks, and J. D. Jorgensen, Phys. Rev. Lett., **87**, 047001 (2001).

[41] N. R. Werthamer, E. Helfand, and P. C. Hohenberg, Phys. Rev., **147**, 295 (1966).

[42] L. Lyard, P. Szabó, T. Klein, J. Marcus, C. Marcenat, K. H. Kim, B. W. Kang, H. S. Lee, and S. I. Lee, Phys. Rev. Lett., **92**, 057001 (2004).

[43] S. Manalo, H. Michor, M. El-Hagary, G. Hilscher, and E. Schachinger, Phys. Rev. B, **63**, 104508 (2001).

[44] S. V. Shulga, S.-L. Drechsler, G. Fuchs, K.-H. Müller, K. Winzer, M. Heinecke, and K. Krug, Phys. Rev. Lett., **80**, 1730 (1998).

[45] R. Goyal, R. Jha, B. Tiwari, A. Dixit, and V. P. S. Awana, Supercond. Sci. Technol., **29**, 075008 (2016).

[46] P. Cheng, B. Shen, F. Han and H.-H. Wen, Europhysics Lett., **104**, 37007 (2013).

[47] S. J. Singh, J. Shimoyama, A. Yamamoto, H. Ogino, and K. Kishio, Physica C, **494**, 57 (2013).

[48] Anupam, V. K. Anand, P. L. Paulose, S. Ramakrishnan, C. Geibel, and Z. Hossain, Phys. Rev. B, **85**, 144513 (2012).

[49] A. A. Golubov, and I. I. Mazin, Phys. Rev. B, **55**, 15146 (1997).

[50] Y. Yamakawa, S. Onari, and H Kontani, Phys. Rev. B, **87**, 195121 (2013).

[51] T. Kawamata, E. Satomi, Y. Kobayashi, M. Itoh, and M. Sato, J. Phys. Soc. Jpn., **80**, 084720 (2011).